\documentclass[a4paper,conference]{IEEEtran}
\usepackage{cite}
\usepackage{xcolor} %

\ifCLASSINFOpdf
  \usepackage[pdftex]{graphicx}
\else
\fi
\usepackage{amsmath}
\usepackage{array}

\usepackage[caption=false,font=normalsize,labelfont=sf,textfont=sf]{subfig}
\usepackage{fixltx2e}
\usepackage{eso-pic}
\newcommand\AtPageUpperMyright[1]{\AtPageUpperLeft{%
 \put(\LenToUnit{0.2\paperwidth},\LenToUnit{-1cm}){%
     \parbox{0.8\textwidth}{\raggedleft\fontsize{9}{11}\selectfont #1}}%
 }}%
\newcommand{\conf}[1]{%
\AddToShipoutPictureBG*{%
\AtPageUpperMyright{#1}
}
}

\begin{document}

\title{3D Medical Multi-modal Segmentation Network Guided by Multi-source Correlation Constraint}



\author{\IEEEauthorblockN{Tongxue Zhou\IEEEauthorrefmark{1}\IEEEauthorrefmark{2}\IEEEauthorrefmark{3},
St\'ephane Canu\IEEEauthorrefmark{1}\IEEEauthorrefmark{3},
Pierre Vera\IEEEauthorrefmark{4} and
Su Ruan\IEEEauthorrefmark{2}\IEEEauthorrefmark{3}}
\IEEEauthorblockA{\IEEEauthorrefmark{1}INSA Rouen, LITIS - Apprentissage, Rouen 76800, France\\}
\IEEEauthorblockA{\IEEEauthorrefmark{2}Universit\'e de Rouen Normandie, LITIS - QuantIF, Rouen 76183, France\\}
\IEEEauthorblockA{\IEEEauthorrefmark{3}Normandie Univ, INSA Rouen, UNIROUEN, UNIHAVRE, LITIS, France\\}
\IEEEauthorblockA{\IEEEauthorrefmark{4} Department of Nuclear Medicine, Henri Becquerel Cancer Center, Rouen, 76038, France}}


\maketitle
\begin{flushleft}
{\conf{\large\textcolor{red}{Accepted in 25th International Conference on Pattern Recognition (ICPR)\\Jan 2021, Milan, Italy}}}
\end{flushleft}
\vspace{1cm}
\begin{abstract}
In the field of multimodal segmentation, the correlation between different modalities can be considered for improving the segmentation results. In this paper, we propose a multi-modality segmentation network with a correlation constraint. Our network includes N model-independent encoding paths with N image sources, a correlation constrain block, a feature fusion block, and a decoding path. The model independent encoding path can capture modality-specific features from the N modalities. Since there exists a strong correlation between different modalities, we first propose a linear correlation block to learn the correlation between modalities, then a loss function is used to guide the network to learn the correlated features based on the linear correlation block. This block forces the network to learn the latent correlated features which are more relevant for segmentation. Considering that not all the features extracted from the encoders are useful for segmentation, we propose to use dual attention based fusion block to recalibrate the features along the modality and spatial paths, which can suppress less informative features and emphasize the useful ones. The fused feature representation is finally projected by the decoder to obtain the segmentation result. Our experiment results tested on BraTS-2018 dataset for brain tumor segmentation demonstrate the effectiveness of our proposed method.

\end{abstract}


%
\IEEEpeerreviewmaketitle

\section{Introduction}

Multimodal segmentation using a single model remains challenging due to the different image characteristics of different modalities. A key challenge is to exploit the latent correlation between modalities and to fuse the complementary information to improve the segmentation performance. In this paper, we proposed a method to exploit the multi-source correlation and apply it to brain tumor segmentation task.

A brain tumor is a growth of cells in the brain that multiplies in an abnormal, uncontrollable way, which is one of the most lethal cancers in the world. Today, an estimated 700,000 people in the United States are living with a primary brain tumor, and over 87,000 more will be diagnosed in 2020\footnote{NBTS: National Brain Tumor Society}. Gliomas are the most common brain tumors that arise from glial cells. According to the malignant degree of gliomas \cite{menze2014multimodal}, they can be categorized into two grades: low-grade gliomas (LGG) and high-grade gliomas (HGG), the former one tend to be benign, grow more slowly with lower degrees of cell infiltration and proliferation, the latter one are malignant, more aggressive and need immediate treatment, moreover, the five-year relative survival rate of gliomas is only 6.8\%. Therefore, early diagnosis of brain tumors is highly desired in clinical practice for better treatment planning.

Magnetic Resonance Imaging (MRI) is commonly used in radiology to diagnose brain tumors, it is a non-invasive and good soft tissue contrast imaging modality, which provides invaluable information about shape, size, and localization of brain tumors without exposing the patient to a high ionization radiation \cite{liang2000principles,bauer2013survey,drevelegas2010imaging}. The commonly used sequences are T1-weighted (T1), contrast-enhanced T1-weighted (T1c), T2-weighted (T2) and Fluid Attenuation Inversion Recovery (FLAIR) images. In this  work, we refer to these images of different sequences as modalities. Different modalities can provide complementary information to analyze different sub-regions of gliomas. For example, T2 and FLAIR highlight the tumor with peritumoral edema, designated whole tumor. T1 and T1c highlight the tumor without peritumoral edema, designated tumor core. An enhancing region of the tumor core with hyper-intensity can also be observed in T1c, designated enhancing tumor core. Therefore applying multi-modal images can reduce the information uncertainty and improve clinical diagnosis and segmentation accuracy. 

Inspired by a fact that, there is strong correlation between multi MR modalities, since the same scene (the same patient) is observed by different modalities \cite{lapuyade2017segmenting}. We propose a 3D multimodal brain segmentation network guided by multi-source correlation constrain. The main contributions of our method are four folds: 1) A correlation block is introduced to discover the latent multi-source correlation between modalities, making the features more relevant for segmentation. 2) A dual attention based fusion strategy is proposed to recalibrate the feature representation along modality-wise and spatial-wise. 3) A correlation based loss function is proposed to aide the segmentation network to extract the correlated feature representation for a better segmentation. 4) The first 3D multimodal brain tumor segmentation network guided by multi-source correlation constrain is proposed. 

\subsection{Related Work}
\begin{figure}[htb]
\centering
\includegraphics[width=3.5in]{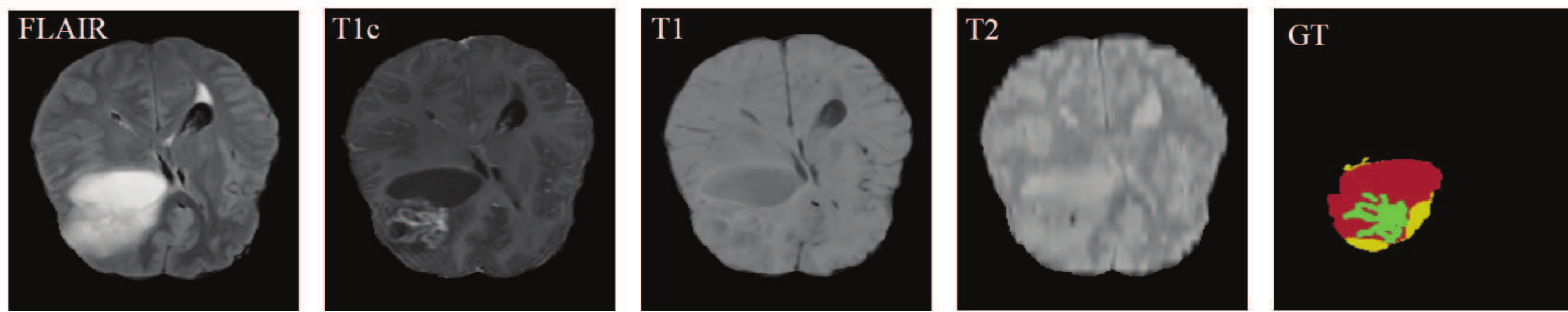}
\caption{Example of data from a training subject. The first four images from left to right show the MRI modalities: Fluid Attenuation Inversion Recovery (FLAIR), contrast enhanced T1-weighted (T1c), T1-weighted (T1), T2-weighted (T2) images, and the fifth image is the ground truth labels created by experts. The color is used to distinguish the different tumor regions: red: Necrotic and Non-enhancing tumor, yellow: edema, green: enhancing tumor, black: healthy tissue
and background.}
\label{fig0}
\end{figure}

A wide range of approaches for brain tumor segmentation, such as probability theory \cite{lapuyade2017segmenting}, kernel feature selection \cite{zhang2011kernel}, belief function \cite{lian2018joint} based on \cite{lian2016dissimilarity}, random forests \cite{zikic2012decision}, conditional random fields \cite{yu2018semi} and support vector machines \cite{bauer2011fully} have been developed with success. However, brain tumor segmentation is still a challenging task due to three reasons: (1) The brain anatomy structure varies from patients to patients. (2) The variability across size, shape, and texture of gliomas. (3) The variability in intensity range and low contrast in qualitative MR imaging modalities (see Fig.~\ref{fig0}).

Recently, with a strong feature learning ability, deep learning-based approaches have become more prominent for brain tumor segmentation. Cui et al. \cite{cui2018automatic} proposed a cascaded deep learning convolutional neural network consisting of two sub-networks. The first network is to define the tumor region from a MRI slice and the second network is used to label the defined tumor region into multiple sub-regions. Zhao et al. \cite{zhao2018deep} integrated fully convolutional neural networks (FCNNs) \cite{long2015fully} and conditional random fields to segment brain tumor. Havaei et al. \cite{havaei2017brain} implemented a two-pathway architecture that learns about the local details of the brain tumor as well as the larger context feature. Wang et al. \cite{wang2017automatic} proposed to decompose the multi-class segmentation problem into a sequence of three binary segmentation problems according to the sub-region hierarchy. Kamnitsas et al. \cite{kamnitsas2017efficient} proposed an efficient fully connected multi-scale CNN architecture named DeepMedic, which reassembles a high resolution and a low resolution pathway to obtain the segmentation results. Furthermore, they used a 3D fully connected conditional random field to effectively remove false positives. Kamnitsas et al. \cite{kamnitsas2017ensembles} introduced EMMA, an ensemble of multiple models and architectures including DeepMedic, FCNs and U-Net. Myronenko et al. \cite{myronenko20183d} proposed a segmentation network for brain tumor from multimodal 3D MRIs, where variational auto-encoder branch is added into the U-net to further regularize the decoder in the presence of limited training data.

For multimodal segmentation task, exploiting the complimentary information from different modalities play an import role in the final segmentation accuracy. As presented in \cite{zhou2019review}, the multi-modal segmentation network architectures can be categorized into single-encoder-based method and multi-encoder-based method. The single-encoder-based method~\cite{isensee2017brain,kamnitsas2017ensembles} directly integrates the different multi-modality images channel-wise in the input space, while the correlations between different modalities are not well exploited. However, the multi-encoder-based method \cite{tseng2017joint}, 
allows to separately extract individual feature information by applying multiple modality-specific encoders, and to fuse them with specific fusion strategy to emphasize the useful information for the segmentation task. According to \cite{zhou2020multi}, multi-encoder-based method has better performance than single-encoder-based method, which can learn more complementary and cross-modal interdependent features. However, not all features extracted from the encoder are useful for segmentation. Therefore, it is necessary to find an effective way to fuse features, we focus on the extraction of the most informative features for segmentation. To this end, we propose to use the attention mechanism, which can be viewed as a tool being capable to take into account the most informative feature representation. 
Channel attention modules and spatial attention modules are the commonly used attention mechanisms. The former one learn a channel-wise feature representation that quantifies the relative importance
of each channel’s features \cite{hu2018squeeze, li2018pyramid, oktay2018attention}. The latter one, spatial attention modules, learn the feature representation in each position by weighted sum the features of all other positions \cite{roy2018concurrent,roy2020squeeze,fu2019dual}. However, the methods mentioned above evaluated the attention mechanism only on the single-modal image dataset and don’t consider the fusion issue on the multi-modal medical images. In this paper, we propose to apply the attention mechanism on the multi-modality brain tumor dataset. To learn the contributions of the feature representations from different modalities, we propose a dual attention based fusion block to selectively emphasize feature representations, which consists of a modality attention module and a spatial attention module. The proposed fusion block uses the modality-specific features to derive a modality-wise and a spatial-wise weight map that quantify the relative importance of each modality’s features and also of the different spatial locations in each modality. These fusion maps are then multiplied with the modality-specific feature representations to obtain a fused representation of the complementary multi-modality information. In this way, we can discover the most relevant characteristics to aide the segmentation.

For multi-modal MR brain tumor segmentation, since the four MR modalities are from the same patient, there exists a strong correlation between modalities \cite{lapuyade2017segmenting}. In this paper, our goal is to exploit and utilize the correlation between modalities to improve the segmentation performance. Therefore, we first exploit the correlation between each two modalities and then utilize a loss function to guide the segmentation network to learn the correlated features to enhance the segmentation result. To the best of our knowledge, this is the first work which is capable of utilizing the latent multi-source correlation to help the segmentation.



\section{Method}
\begin{figure*}[htb]
\centering
\includegraphics[width=7in]{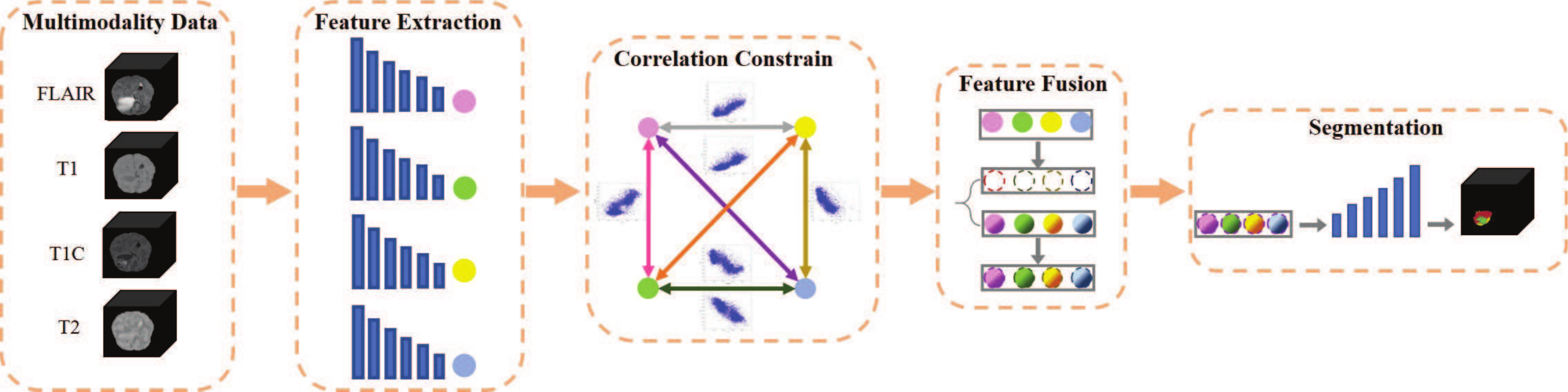}
\caption{The pipeline of the proposed method, consisting of feature extraction, correlation constrain and fusion fusion block, 4 color circles represent 4 modality feature representations.}
\label{fig1}
\end{figure*}

Our network is based on our previous work \cite{zhou2020multi}, which used a multi-encoder based network to deal with the multi-model fusion issue. In this paper, we aim to exploit the multi-source correlation between modalities and utilize the correlation to constrain the network to learn more effective feature so as to improve the segmentation performance. To learn complementary features and cross-modal inter-dependencies from multi-modality MRIs, we applied the multi-encoder based framework. It takes 3D MRI modality as input in each encoder. Each encoder can produce a modality-specific feature representation, at the lowest level of the network, the linear correlation block is first used to exploit the latent multi-source correlation, then a well-designed loss function is applied to guide the network to learn the effective feature information. Then, all the modality-specific feature representations are concatenated to the fusion block at each layer. With the assistance of the dual attention fusion block, the feature representations will be separated along modality-wise and space-wise, and the most informative feature is obtained as the shared latent representation, and finally it is projected by decoder to the label space to obtain the segmentation result. The pipeline of our method is described in Fig.~\ref{fig1}.

\begin{figure*}[htb]
\centering
\includegraphics[width=7in]{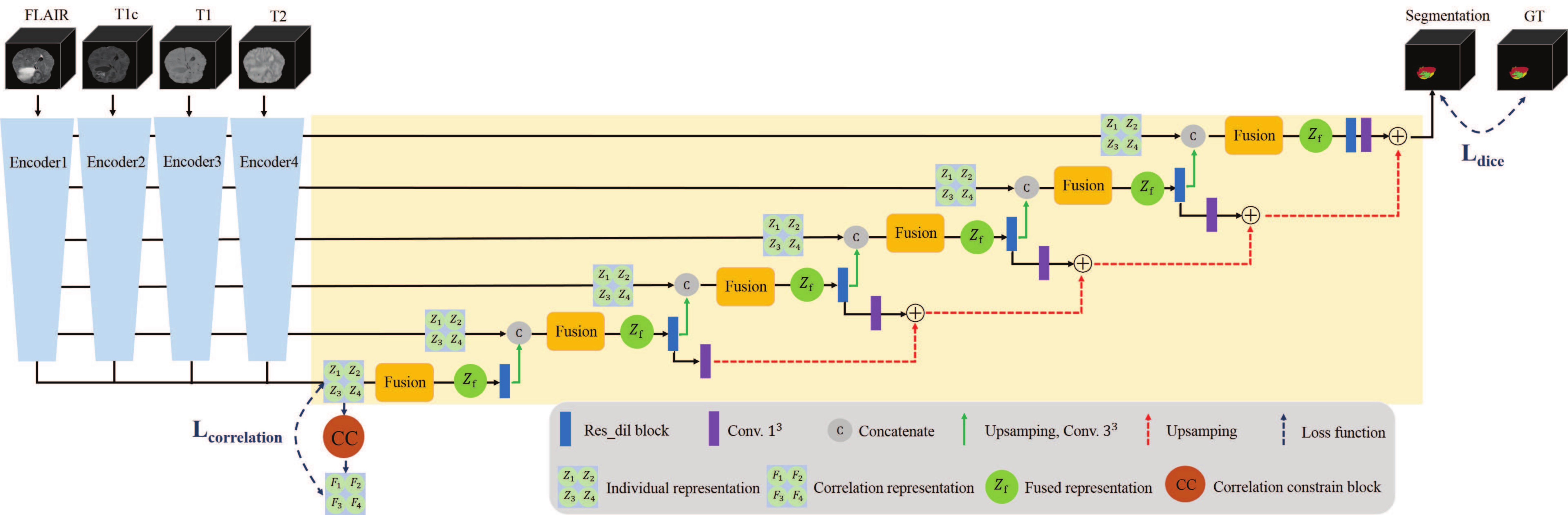}
\caption{Overview of our proposed segmentation network framework.}
\label{fig2}
\end{figure*}

\subsection{Architecture Design}
It's likely to require different receptive fields when segmenting different regions in an image, a standard U-Net can’t get enough semantic features due to the limited receptive field. Inspired by dilated convolution, we use residual block with dilated convolutions (rate = 2, 4) (res\_dil block) on both encoder part and decoder part to obtain features at multiple scale. The encoder includes a convolutional block, a res\_dil block followed by skip connection. All convolutions are $3\times3\times3$. Each decoder level begins with up-sampling layer followed by a convolution to reduce the number of features by a factor of 2. Then the upsampled features are combined with the features from the corresponding level of the encoder part using concatenation. After the concatenation, we use the res\_dil block to increase the receptive field. In addition, we employ deep supervision~\cite{isensee2017brain} for the segmentation decoder by integrating segmentation layers from different levels to form the final network output. The proposed network architecture is described in Fig.~\ref{fig2}.

\subsection{Correlation Constrain Block}

\begin{figure}[!t]
\centering
\subfloat[]{\includegraphics[width=1in]{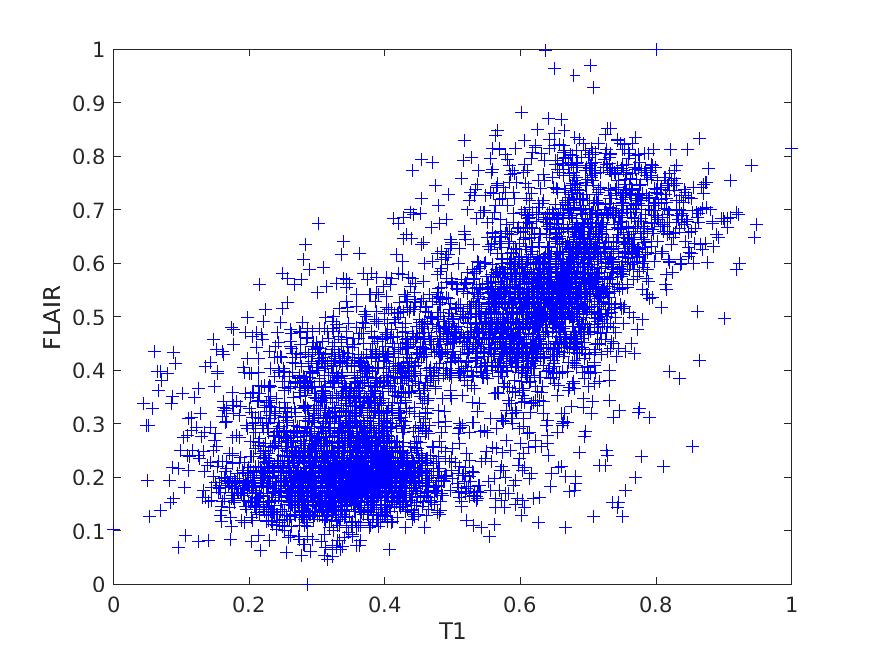}%
\label{a}}
\hfil
\subfloat[]{\includegraphics[width=1in]{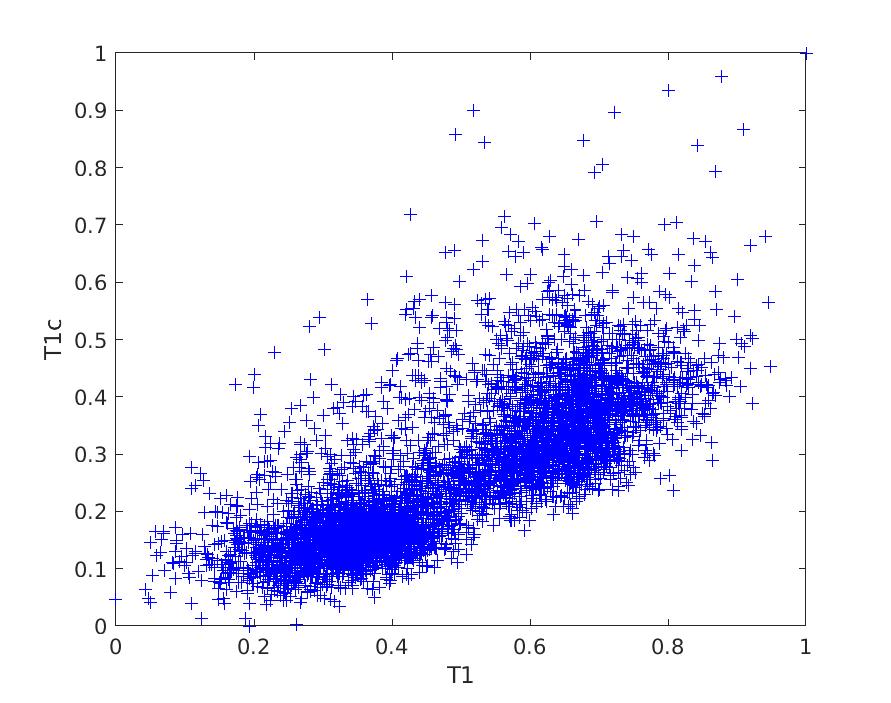}%
\label{b}}
\hfil
\subfloat[]{\includegraphics[width=1in]{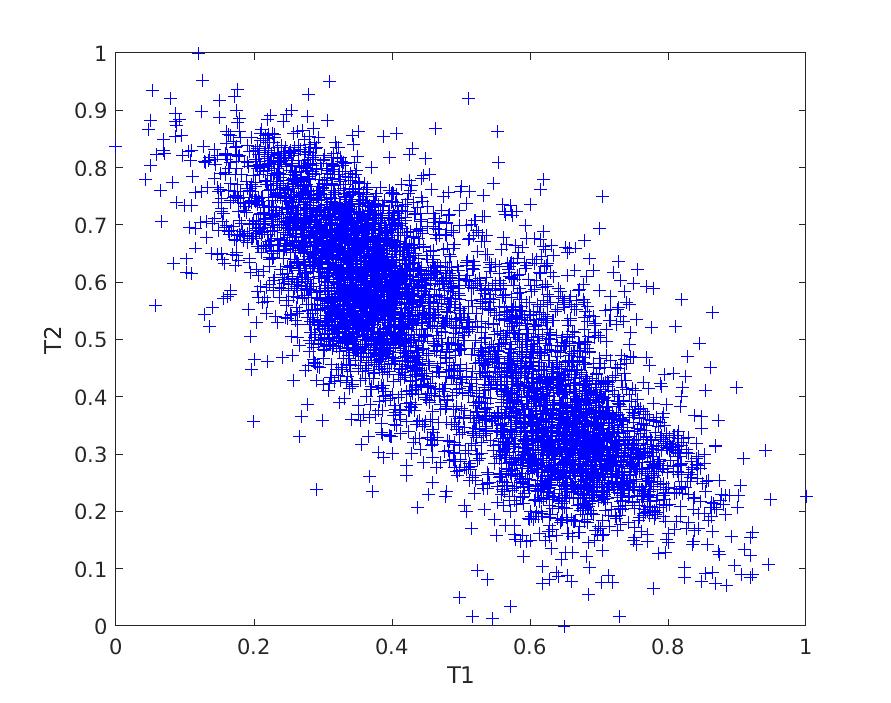}%
\label{c}}
\hfil
\subfloat[]{\includegraphics[width=1in]{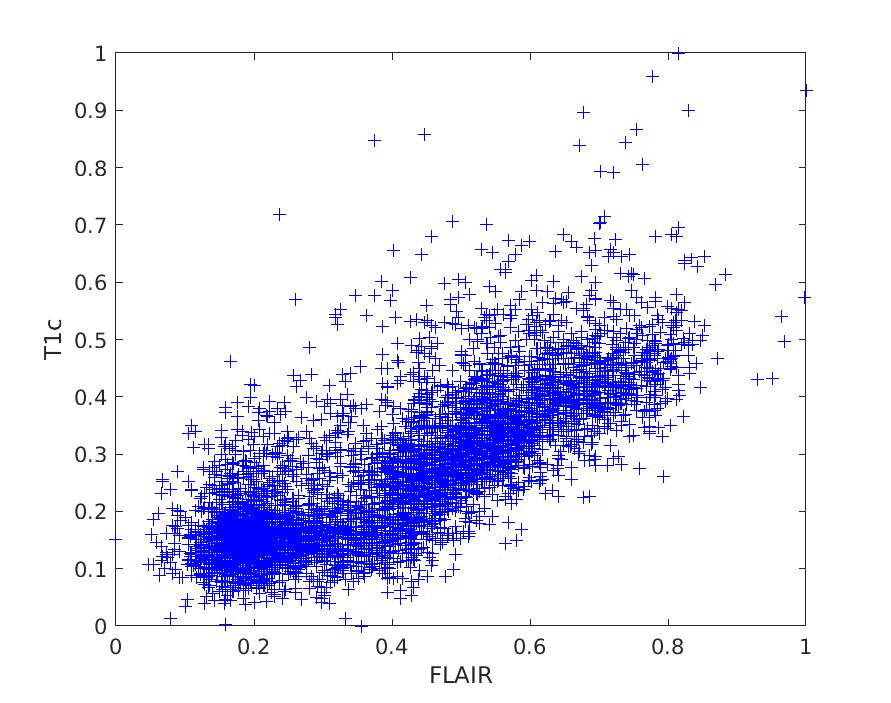}%
\label{d}}
\hfil
\subfloat[]{\includegraphics[width=1in]{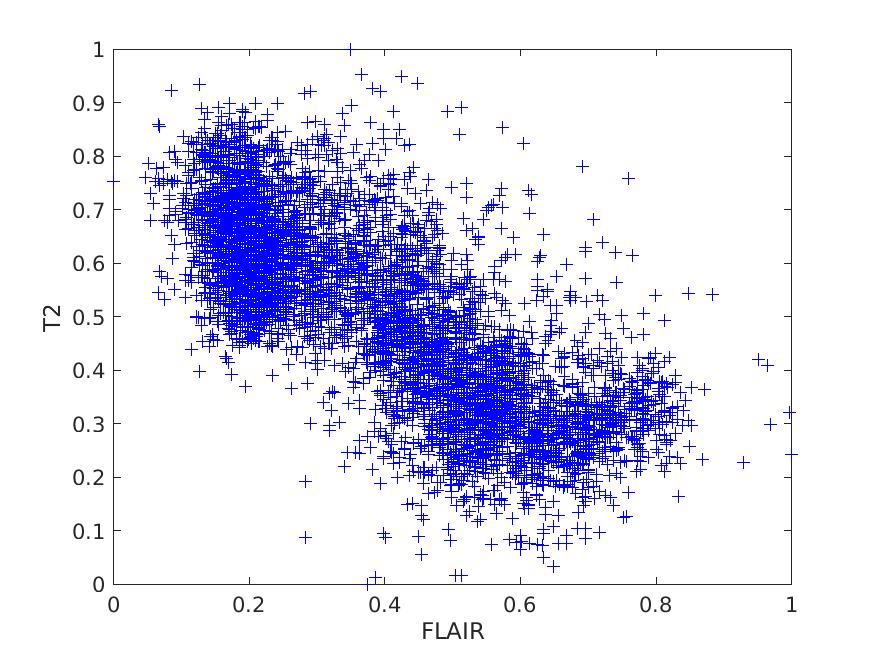}%
\label{e}}
\hfil
\subfloat[]{\includegraphics[width=1in]{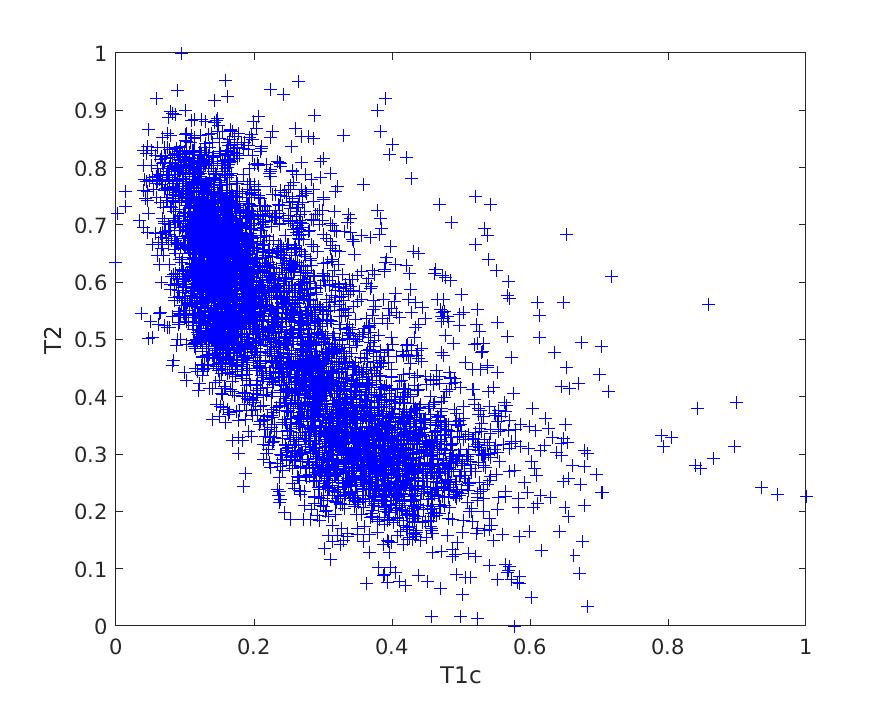}%
\label{f}}
\caption{Joint intensity distributions of MR images: (a) T1-FLAIR, (b) T1-T1c,(c)  T1-T2, (d) FLAIR-T1c,  (e) FLAIR-T2, (f)  T1c-T2. The intensity of the first modality is read on abscissa axis and that of the second modality on the ordinate axis.}
\label{fig3}
\end{figure}

\begin{figure}[htb]
\centering
\includegraphics[width=2.5in]{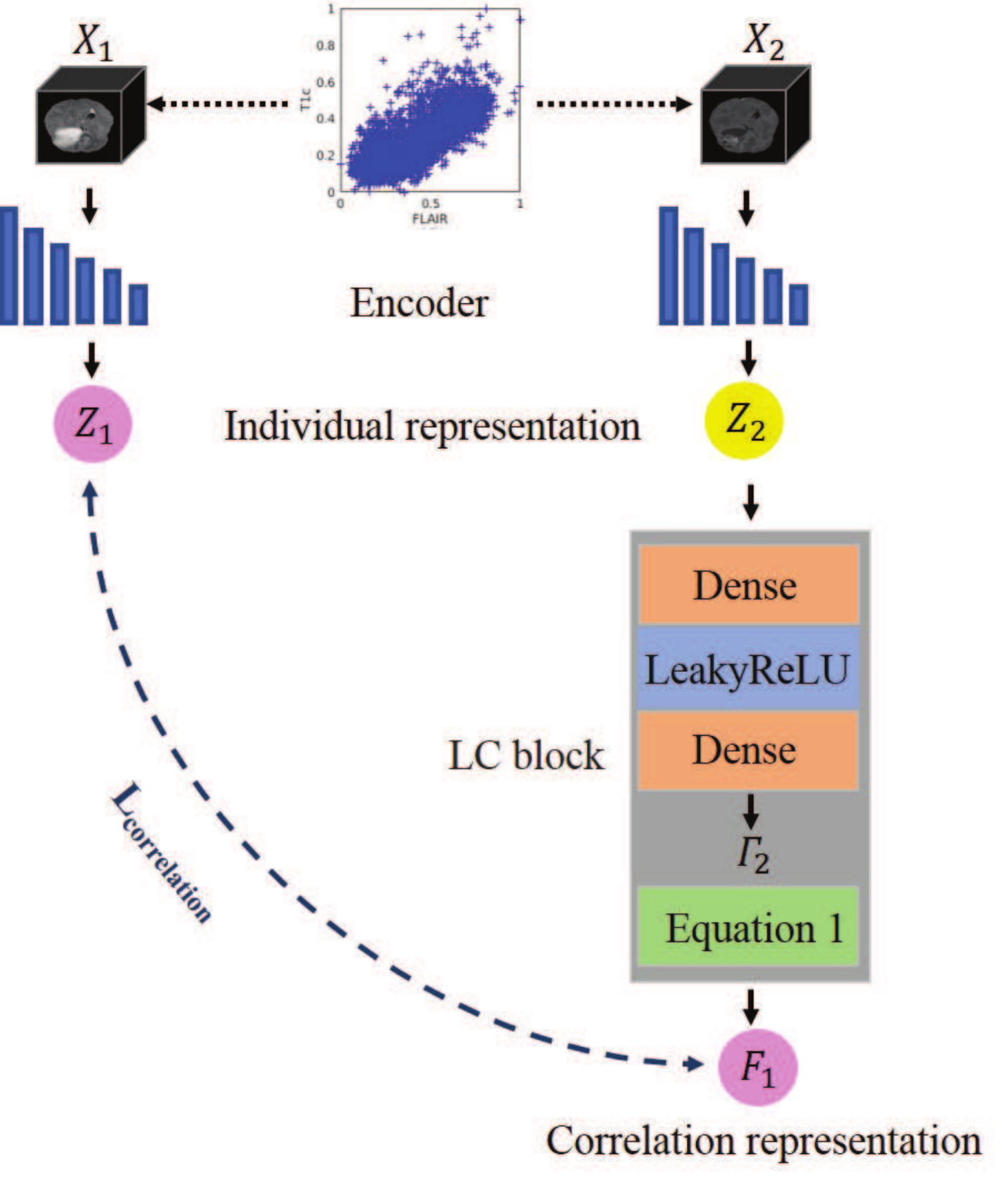}  
\caption{Architecture of Correlation Constrain (CC) block, which consists of Linear Correlation (LC) block and a correlation constrain loss.}
\label{fig4}
\end{figure}

Inspired by a fact that, there is strong correlation between multi MR modalities, since the same scene (the same patient) is observed by different modalities~\cite{lapuyade2017segmenting}. From Fig.~\ref{fig3} presenting joint intensities of the MR images, we can observe a strong correlation in intensity distribution between each two modalities. To this end, it's reasonable to assume that a strong correlation also exists in latent representation between modalities. Therefore, we introduce a Correlation Constrain (CC) block, which consists of a Linear Correlation (LC) block (see Fig.~\ref{fig4}) to discover the latent correlation and a correlation loss to constrain the correlation between modalities. For simplicity, we present the CC block using two modalities. The input modality $\{X_i, ... , X_n\}$, where $n = 4$, is first input to the independent encoders (with learning parameters $\theta$) to learn the modality-specific representation $Z_i(X_i|\theta_i)$. Then, a network with two fully connected network with LeakyReLU, maps the modality-specific representation $Z_i(X_i|\theta_i)$ to a set of independent parameters $\Gamma_i =\{\alpha_i, \beta_i\}$, $i=1,... ,n$. Finally the linear correlation representation of $j$ modality  $F_j(X_j|\theta_j)$ can be obtained via linear correlation Equation~\ref{eq1}. 


Since we have four modalities, and each two modalities have a strong linear correlation, we only need to learn three pairs of correlation expressions from each two modalities. Then, the Kullback–Leibler divergence (Equation~\ref{eq2}) is used as the correlation loss to constrain the distributions between the estimated correlation representation and the original feature representation, which enables the segmentation network to learn the correlated feature representation to improve the segmentation performance.

\begin{equation}
    F_j(X_j|\theta_j) = \alpha_i \odot Z_i(X_{i}|\theta_{i})+\beta_i,
    (i \neq j)
\label{eq1}
\end{equation}

\begin{equation}
 L_{correlation} = \sum_{x\in X} P(x)log\frac{P(x)}{Q(x)}
 \label{eq2}
\end{equation}

\noindent where $P$ and $Q$ are probability distributions of $Z_i$ and $F_j$, respectively, which defined on the same probability space $X$.


\subsection{Dual Attention based fusion strategy}
\begin{figure}[htb]
\centering
\includegraphics[width=3.5in]{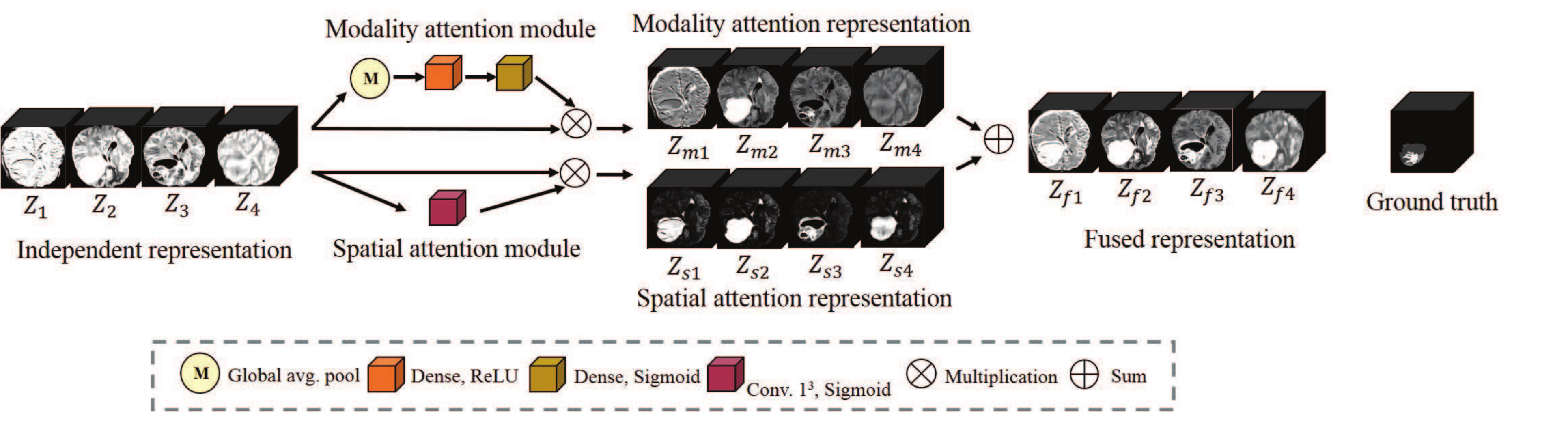} 
\caption{Proposed dual attention fusion block. The individual feature  representations ($Z_1$, $Z_2$, $Z_3$, $Z_4$) are first concatenated, then they are recalibrated along modality attention module and spatial attention module to achieve the modality attention representation $Z_m$ and spatial attention representation $Z_s$, final they are added to obtain the fused feature representation $Z_f$.}
\label{fig5}
\end{figure}

The purpose of fusion is to stand out the most important features from different source images to highlight regions that are greatly relevant to the target region. Since different MR modalities can identify different attributes of the target tumor to provide complementary information. In addition, from the same MR modality, we can learn different content at different locations. Inspired by the attention mechanism \cite{roy2018concurrent}, we propose a dual attention based fusion block to enable a better integration of the complementary information between modalities, which consists of a modality attention module and a spatial attention module, the architecture is described in Fig.~\ref{fig5}.
 
The individual feature representations learned by four encoders ($Z_1$, $Z_2$, $Z_3$, $Z_4$) are first concatenated to obtain the input feature representation $Z= [Z_1, Z_2, Z_3, Z_4]$, $Z_k\in R^{H\times W}$. Note that, in the lowest level of the network, there are four modality-specific feature representations ($Z_1$, $Z_2$, $Z_3$, $Z_4$), in the other levels, the upsamlping layer in the decoder path is also concatenated with the modality-specific feature representations to obtain the input feature representation $Z= [Z_1, Z_2, Z_3, Z_4, Z_5]$, $Z_k\in R^{H\times W}$, for simplicity, in the following, we describe the fusion block with the four modality-specific feature representations. 

In the modality attention module, a global average pooling is first performed to produce a tensor $g\in R^{1\times1\times 4}$, which represents the global spatial information of the feature representation, with its $k^{th}$ element
\begin{equation}
    g_k= \frac{1}{H\times W}\sum_i^H\sum_j^W Z_k(i,j)
\end{equation}
Then two fully-connected layers are applied to encode the modality-wise dependencies, $\hat{g}= W_1(\delta(W_2 g))$, with $W_1\in R^{4\times2}$, $W_2\in R^{2\times4}$, being weights of two fully-connected layers and the ReLU operator $\delta(\cdot)$, $\hat{g}$ is then passed through the sigmoid layer to obtain the modality-wise weights, which will be applied to the input representation $Z$ through multiplication to achieve the modality-wise features $Z_m$, and the $\sigma(\hat{g_k})$ indicates the importance of the $i$ modality of the feature representation.
\begin{equation}
    Z_m= [\sigma (\hat{g_1})Z_1, \sigma(\hat{g_2})Z_2, \sigma(\hat{g_3})Z_3, \sigma(\hat{g_4})Z_4,]
\end{equation}

In the spatial attention module, the feature representation can be considered as $Z= [Z^{1,1}, Z^{1,2}, ... , Z^{i,j},..., Z^{H,W}]$, $Z^{i,j}\in R^{1\times1\times4}$, $i\in{1, 2,..., H}$, $j\in{1, 2,...,W}$, and then a convolution operation $q=W_s\star Z$, $q\in R^{H\times W}$  with weight $W_s\in R^{1\times1\times4\times1}$, is used to squeeze the spatial domain, and to produce a projection tensor, which represents the linearly combined representation for all modalities for a spatial location. The tensor is finally passed through a sigmoid layer to obtain the space-wise weights, $\sigma(q_{i,j})$ indicates the importance of the spatial information $(i, j)$ of the feature representation.

\begin{equation}
    Z_s=[\sigma(q_{1,1})Z^{1,1},  ... , \sigma(q_{i,j})Z^{i,j}, ... , \sigma(q_{H,W})Z^{H,W}]
\end{equation}

Finally, the learned fused feature representation is obtained by adding the modality-wise feature representation and space-wise feature representation.

\begin{equation}
    Z_f=Z_m+Z_s 
\end{equation}

From Fig.~\ref{fig5}, we can observe the target tumor's characteristics in the four independent feature representations are not obvious, however, the modality attention module stands out the different attributes of the modalities to provide complementary information, for example, the FLAIR modality highlights the edema region and T1c modality stand out the tumor core region. In the spatial attention module, all the locations related to the target tumor region are highlighted, In this way, we can discover the most relevant characteristics between modalities. Furthermore, the proposed fusion block can be directly adapted to any multi modal fusion problem. 

\section{Data and Implementation Details}
\subsection{Data}
The datasets used in the experiments come from BraTS  2018  dataset. The  training  set  includes  285  patients,  each  patient  has four image modalities including T1, T1c, T2 and FLAIR. Following the challenge, four intra-tumor structures have been grouped into three mutually inclusive tumor regions: (a) whole tumor (WT) that consists of all tumor tissues, (b) tumor core (TC) that consists of the enhancing tumor, necrotic and non-enhancing tumor core, and (c) enhancing tumor (ET). The provided data have been pre-processed by organisers: co-registered to the same anatomical template, interpolated to the same resolution ($1mm^3$) and skull-stripped. The ground truth have been manually labeled by experts. We did additional pre-processing with a standard procedure. The N4ITK \cite{avants2009advanced} method is used to correct the distortion of MRI data, and intensity normalization is applied to normalize each  modality of each patient. To exploit the spatial contextual information of the image, we use 3D image, crop and resize it from $155\times240\times240$ to $128\times128\times128$.

\subsection{Implementation Details}
Our network is implemented in Keras with a single Nvidia GPU Quadro P5000 (16G). The models are optimized using the Adam optimizer(initial learning rate  =  5e-4) with a decreasing learning rate factor 0.5 with patience of 10 epochs, to avoid over-fitting, early stopping is used when the validation loss isn't improved for 50 epoch. We randomly split the dataset into 80\% training and 20\% testing.  

\subsection{The choices of loss function}
The network is trained by the overall loss function as follow:

\begin{equation}
    \ L_{total}= L_{dice} + \lambda L_{correlation}
\end{equation}

\noindent where $\lambda$ is the trade-off parameters weightig the importance of each component, which is set as 0.1 in our experiment. 

For segmentation, we use dice loss to evaluate the overlap rate of prediction results and ground truth.

\begin{equation}
    \ L_{dice}=1-2\frac{\sum_{i=1}^C\sum_{j=1}^N p_{ic} g_{ic}+\epsilon} {\sum_{i=1}^C\sum_{j=1}^N p_{ic} + g_{ic}+\epsilon}
\end{equation}

\noindent where $N$ is the set of all examples, $C$ is the set of the classes, $p_{ic}$ is the probability that pixel $i$ is of the tumor class $c$ and $p_{i\overline{c}}$ is the probability that pixel $i$ is of the non-tumor class $\overline{c}$. The same is true for $g_{ic}$ and $g_{i\overline{c}}$, and $\epsilon$ is a small constant to avoid dividing by 0.

\subsection{Evaluation metrics}
To evaluate the proposed method, two evaluation metrics: Dice Score and Hausdorff distance are used to obtain quantitative measurements of the segmentation accuracy:

\noindent 1) Dice Score: It is designed to evaluate the overlap rate of prediction results and ground truth. It ranges from 0 to 1, and the better predict result will have a larger Dice value.

\begin{equation}
 Dice = \frac {2TP}{2TP+FP+FN}
\end{equation}

\noindent where $TP$ represents the number of true positive voxels, $FP$ represents the number of false positive voxels, and $FN$ represents the number of false negative voxels. 

\noindent 2) Hausdorff distance (HD): It is computed between boundaries of the prediction results and ground-truth, it is an indicator of the largest segmentation error. The better predict result will have a smaller HD value.

\begin{equation}
  HD=\max\{sup_{r_\in{\partial R}}d_m(s,r),sup_{s_\in\partial S}d_m(r,s)\}
\end{equation}

\noindent where $\partial S$ and $\partial R$ are the sets of tumor border voxels for the predicted and the real annotations, and $d_m(v,v)$ is the minimum of the Euclidean distances between a voxel $v$ and voxels in a set $v$.


\section{Experiment Results}
We conduct a series of comparative experiments to demonstrate the effectiveness of our proposed method and compare it to other approaches. In Section \ref{A}, we first perform an ablation experiment to see the importance of our proposed components and demonstrate that adding the proposed components can enhance the segmentation performance. In Section \ref{B}, we compare our method with the state-of-the-art methods. In Section \ref{4.2}, the qualitative experiment results further demonstrate that our proposed method can achieve a promising segmentation result.

\subsection{Quantitative Analysis}
To prove the effectiveness of our network, we first did an ablation experiment to see the effectiveness of our proposed components, and then we compare our method with the state-of-the-art methods. All the results are obtained by online evaluation platform\footnote{https://ipp.cbica.upenn.edu/}.

\subsubsection{Effectiveness of Individual Modules}
\label{A}
To assess the performance of our method, and see the importance of the proposed components in our network, including dual attention fusion strategy and correlation constrain block, we did an ablation experiment, our network without dual attention fusion strategy and correlation constrain block is denoted as baseline. From Table \ref{tab1}, we can observe the baseline method achieves Dice Score of 0.726, 0.867, 0.766 for enhancing tumor, whole tumor, tumor core, respectively. When the dual attention fusion strategy is applied to the network, we can see an increase of Dice Score and Hausdorff Distance across all tumor regions with an average improvement of 0.85\% and 6.44\%. respectively. The major reason is that the proposed fusion block can help to emphasize the most important representations from the different modalities across different positions in order to boost the segmentation result. In addition, another advantage of our method is using the correlation constrain block, which can constrain the encoders to discover the latent multi-source correlation representation between modalities and then guide the network to learn correlated representation to achieve a better segmentation. From the results, we can observe that with the assistance of correlation constrain block, the network can achieve the best Dice Score of 0.747, 0.886 and 0.776 and Hausdorff Distance of 7.851, 7.345 and 9.016 for enhancing tumor, whole tumor, tumor core, respectively with an average improvement of 2.21\% and 9.28\% relating to the baseline. The results in Table \ref{tab1} demonstrate the effectiveness of each proposed component and our proposed network architecture can perform well on brain tumor segmentation.

\begin{table}[]
\centering
\caption{Evaluation of our proposed method on Brats 2018 training dataset, (1) Baseline (2) Baseline + Dual attention fusion (3) Baseline + Dual attention fusion + Correlation constrain, ET, WT, TC denote enhancing tumor, whole tumor and tumor core, respectively.}
\label{tab1}
\renewcommand{\arraystretch}{1.3}%
\resizebox{0.48\textwidth}{!}{%
\begin{tabular}{ccccccc}
\hline
{Methods} & \multicolumn{3}{c}{Dice Score} & \multicolumn{3}{c}{Hausdorff (mm)} \\ 
& ET & WT & TC & ET & WT & TC \\ \hline
(1) & 0.726 & 0.867 & 0.764 & 8.743 & 8.463 & 9.482 \\
(2) & 0.733 & 0.879 & 0.765 & 8.003 &7.813 & 9.153 \\
(3) & \textbf{0.747} & \textbf{0.886} & \textbf{0.776} & \textbf{7.851} & \textbf{7.345} & \textbf{9.016} \\ \hline
\end{tabular}}
\end{table}

\subsubsection{Comparisons with the State-of-the-art}
\label{B}
To demonstrate the performance of our method, we compare our proposed method with the state-of-the-art methods on Brats 2018 validation set, which contains 66 images of patients without the ground truth. Table \ref{tab2} shows the comparison results. We have also carried out a comparison study with the state of art of methods based on U-Net.




(1) Hu et al. \cite{hu2018brain} proposed the multi-level up-sampling network (MU-Net) for automated segmentation of brain tumors, where a novel global attention (GA) module is used to combine the low level feature maps obtained by the encoder and high level feature maps obtained by the decoder.




(2) Tuan et al. \cite{tuan2018brain} proposed using Bit-plane to generate a series of binary images by determining significant bits. Then, the first U-Net used the significant bits to segment the tumor boundary, and the other U-Net utilized the original images and images with least significant bits to predict the label of all pixel inside the boundary. 

(3) Hu et al. \cite{hu2018hierarchical} introduced the 3D-residual-Unet architecture. The network comprises a context aggregation pathway and a localization pathway, which encoder abstract representation of the input, and then recombines these representations with shallower features to precisely localize the interest domain via a localization path.

(4) Myronenko et al. \cite{myronenko20183d} proposed a 3D MRI brain tumor segmentation using autoencoder regularization, where a variational autoencoder branch is added to reconstruct the input image itself in order to regularize the shared decoder and impose additional constraints on its layers. 

The best result in BraTS 2018 Challenge is from \cite{myronenko20183d}, which achieves 0.814, 0.904 and 0.859 in terms of Dice Score on enhancing tumor, whole tumor and tumor core regions, respectively. However, it uses 32 initial convolution filters and a lot of memories (NVIDIA Tesla V100 32GB GPU is required) to train the model, which is computationally expensive. While our method used only 8 initial filters, and from Table \ref{tab2}, it can be observed that our proposed method can yield a competitive results in terms of Dice Score and Hausdorff distance across all the tumor regions. We also implemented the method \cite{myronenko20183d} with 8 initial filters, but the results are not good. Compared with other methods, \cite{hu2018hierarchical} has a better Dice Score on enhancing tumor, while our method achieves a better average Dice Score on all the tumor regions with an improvement of 1.15\%, and it can also obtain an average improvement of 14.81\% for Hausdorff Distance.

\begin{figure}[htb]
\centering
\includegraphics[width=3.5in]{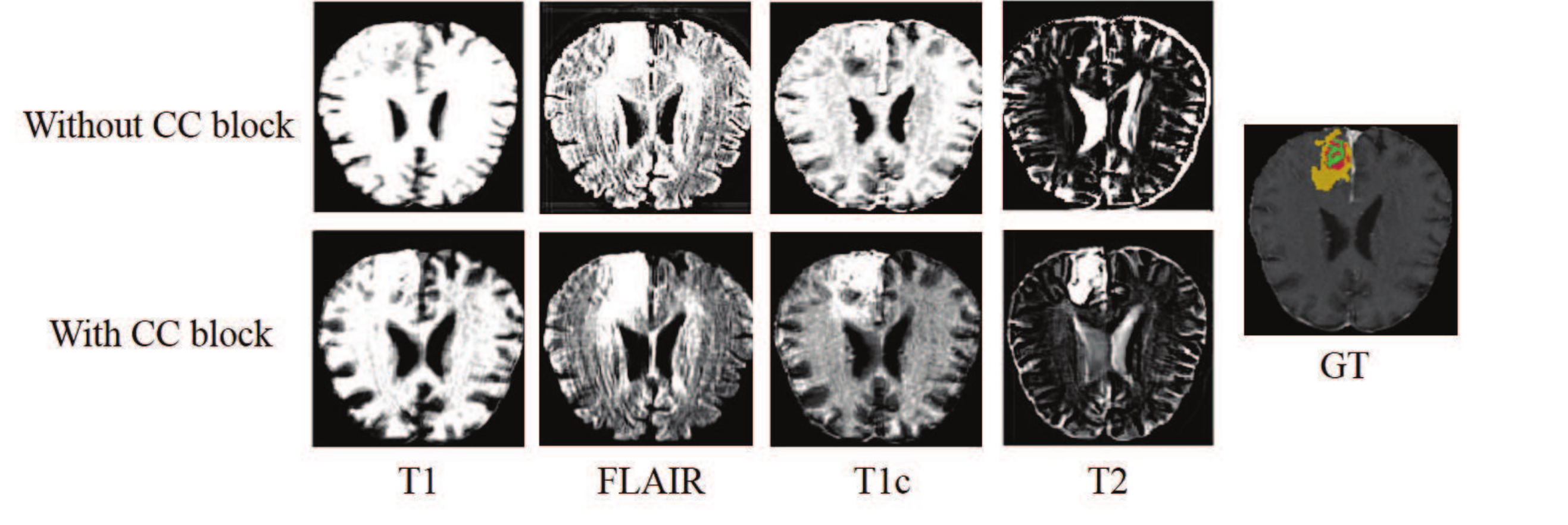}
\caption{Visualization of effectiveness of proposed correlation constrain block.}
\label{fig7}
\end{figure}

To visualize the effectiveness of proposed correlation constrain block, we select an example to show the feature representation of the four modalities in the last layer (before the output) of the network in Fig. \ref{fig7}. The first and second row show the feature representations without and with correlation constrain block, the fifth column shows the ground truth. We can observe that, the correlation constrain block can constrain the network to emphasize the interested tumor region for segmentation.

\begin{table}[]
\centering
\caption{Comparison of different methods on BraTS 2018 validation dataset, ET, WT, TC denote  enhancing tumor, whole tumor, tumor core, respectively, bold results denotes the best score for each tumor region, underline results denotes the second best result.}
\label{tab2}
\renewcommand{\arraystretch}{1.3}%
\resizebox{0.45\textwidth}{!}{%
\begin{tabular}{ccccccccc}
\hline
{Methods} & \multicolumn{4}{c}{Dice Score} & \multicolumn{4}{c}{Hausdorff (mm)} \\ \cline{2-9} 
 & ET & WT & TC & Average & ET & WT & TC & Average \\ \hline
\cite{hu2018brain} & 0.69 & 0.88 & 0.74 & 0.77 & 6.69 & 4.76 & 10.67 &  \underline{7.373} \\
\cite{tuan2018brain} & 0.682 & 0.818 & 0.699 & 0.733 & 7.016 & 9.412 & 12.462 & 9.633 \\
\cite{hu2018hierarchical}& \underline{ 0.719} & 0.856 & 0.769 & 0.781 & \underline{5.5} & 10.843 & \underline{9.985} & 8.776 \\
 \cite{myronenko20183d}& \textbf{0.814} & \textbf{0.904} & \textbf{0.859} & \textbf{0.859} & \textbf{3.804} & \textbf{4.483} & \textbf{8.2777} & \textbf{5.521} \\
Proposed  & 0.705 & \underline{0.883} & \underline{ 0.783} &  \underline{0.79} & 7.27 & \underline{5.111} & 10.047 & 7.476 \\ \hline
\end{tabular}%
}
\end{table}

\subsection{Qualitative Analysis}
\label{4.2}
\begin{figure}[htb]
\centering
\includegraphics[width=3.5in]{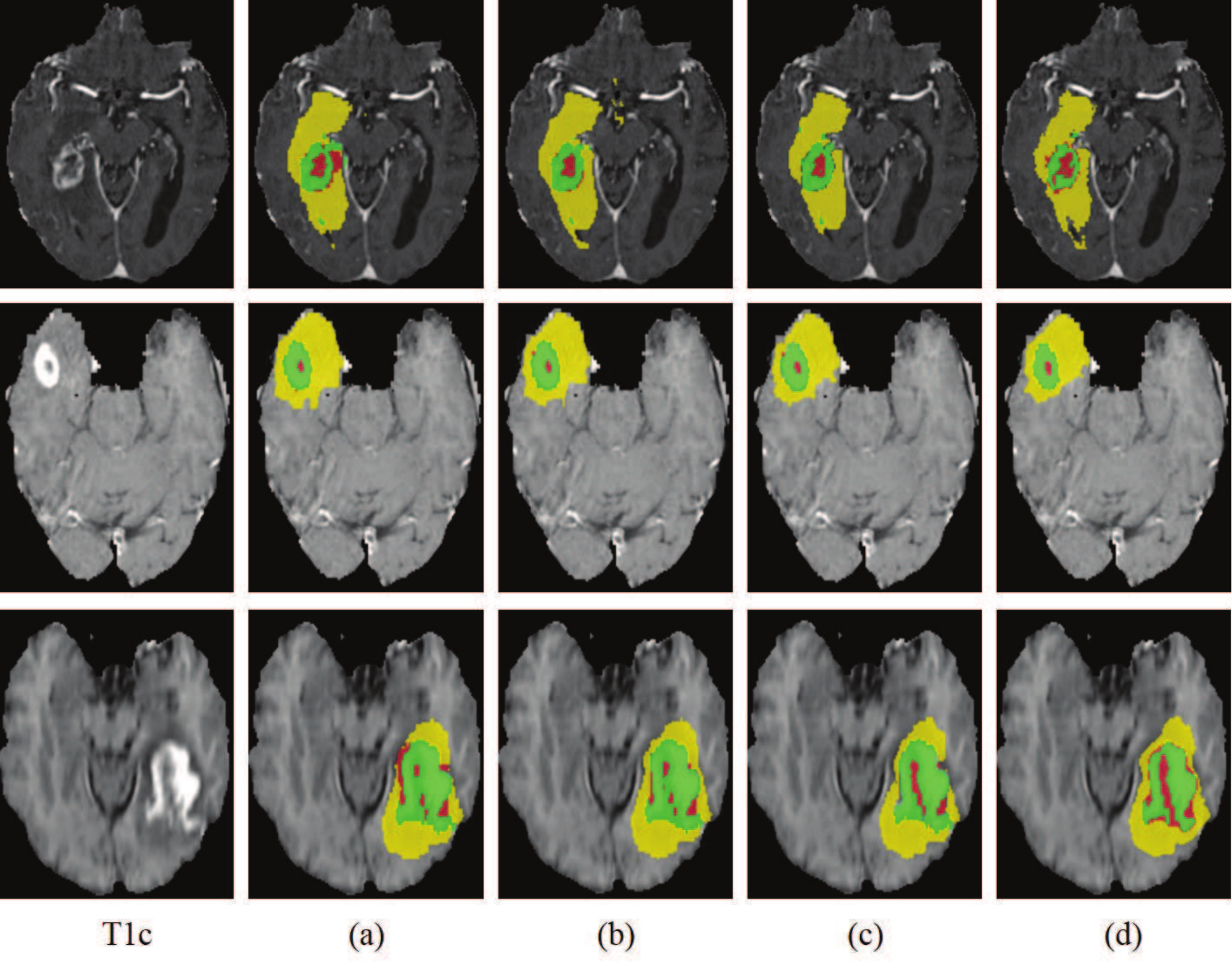}
\caption{Visualization of several segmentation results. (a) Baseline (b) Baseline with fusion block (c) Proposed method with fusion block and correlation constrain (d) Ground truth. Red: necrotic and non-enhancing tumor core; Yellow: edema; Green: enhancing tumor.}
\label{fig8}
\end{figure}

In order to evaluate the robustness of our model, we randomly select several examples on BraTS 2018 dataset and visualize the segmentation results in Fig. \ref{fig8}. From Fig.~\ref{fig8}, we can observe that the segmentation results are gradually improved when the proposed strategies are integrated, these comparisons indicate that the effectiveness of the proposed strategies. In addition, with all the proposed strategies, our proposed method can achieve the best results.

\section{Discussion and Conclusion}
In this paper, we proposed a 3D multimodal brain tumor segmentation network guided by a multi-source correlation constrain, where the architecture demonstrated their segmentation performances in multi-modal MR images of glioma patients. 

To take advantage of the complimentary information from different modalities, the multi-encoder based network is used to learn modality-specific feature representation. Considering the correlation between MR modalities can help the segmentation, a linear correlation block is used to describe the latent multi-source correlation. Since an effective feature learning can contribute to a better segmentation result, a loss function is to guide the network to learn the most correlated feature representation to improve the segmentation. Furthermore, different MR modalities can identify different attributes of the target tumor, and each MR modality image can present different contents at different locations. To this end, inspired by an attention mechanism, a dual-attention fusion strategy is integrated to our network. The modality attention module is used to distinguish the contribution of each modality, and the spatial attention module is used to extract more useful spatial information to boost the segmentation result. The proposed fusion strategy encourages the network to learn more useful feature representation to boost the segmentation result, which is better then the simple max or mean fusion method.

The advantages of our proposed network architecture (i) The segmentation results evaluated on the two metrics (Dice Score and Hausdorff Distance) are similar to real annotation provided by the radiologist experts. (ii) The architecture are an end-to-end Deep Leaning approach and fully automatic without any user interventions. (iii) The experiment results demonstrate that our proposed method gives a very accurate result for the segmentation of brain tumors and its sub-regions even small regions, and it also achieves very competitive results with less computational complexity. In addition, our method can be generalized to other kinds of correlation (e.g. nonlinear) and applied to other kinds of multi-source images if some correlation exists between them.

As a perspective of this research, we will valid our method in different clinical scenarios. In addition, we intend to study a more efficient correlation representation approach to describe the correlation between modalities, and apply it to synthesize additional images to cope with the limited medical image dataset.

\section*{Acknowledgment}
This project was co-financed by the European Union with the European regional development fund (ERDF, 18P03390/18E01750/18P02733) and by the Haute-Normandie Regional Council via the M2SINUM project.



\bibliographystyle{IEEEtran}
\bibliography{IEEEabrv,mybibfile}

%



\end{document}